\documentclass[preprint,aps,prfluids,onecolumn,eqsecnum]{revtex4-2} 

\usepackage{graphicx}
\usepackage{newtxtext}
\usepackage{newtxmath}
\usepackage{natbib}
\usepackage{hyperref}
\usepackage{bm}
\usepackage{cleveref}
\usepackage{subcaption}
\usepackage{color,soul}
\makeatletter
\hypersetup{
    colorlinks = true,
    urlcolor   = blue,
    citecolor  = black,
}
\newcommand{\phantomlabel}[2]{
    \protected@write\@auxout{}{
        \string\newlabel{#2}{
            {\@currentlabel#1}{\thepage}
            {\@currentlabel#1}{#2}{}
        }
    }
    \hypertarget{#2}{}
}
\makeatother
\usepackage{color}

\begin{document}

\begin{abstract}
The squirmer is a popular model to analyse the fluid mechanics of a self-propelled object, such as a micro-organism. We demonstrate that some fore-aft symmetric squirmers can spontaneously self-propel above a critical Reynolds number. Specifically, we numerically study the effects of inertia on spherical squirmers characterised by an axially and fore-aft symmetric `quadrupolar' distribution of surface-slip velocity; under creeping-flow conditions, such squirmers generate a pure stresslet flow, the stresslet sign classifying the squirmer as either a `pusher' or `puller.'  Assuming axial symmetry, and over the examined range of the Reynolds number $Re$ (defined based upon the magnitude of the quadrupolar squirming), we find that spontaneous symmetry breaking occurs in the pusher case above $Re \approx 14.3$, with steady swimming emerging from that threshold consistently with a supercritical pitchfork bifurcation and with the swimming speed growing monotonically with $Re$.
\end{abstract}

\title{Spontaneous locomotion of a symmetric squirmer}

\author{Richard Cobos}
\author{Aditya S. Khair}
\affiliation{Department of Chemical Engineering, Carnegie Mellon University, Pittsburgh, PA 15213, USA}
\author{Ory Schnitzer\footnote[1]{Corresponding author. \textit{Email}: o.schnitzer@imperial.ac.uk}}
\affiliation{Department of Mathematics, Imperial College London, London SW7 2AZ, UK}

\maketitle
\newpage
\section{Introduction}
\label{sec:intro}
The study of the motion of living matter in fluids is a cornerstone of biological fluid mechanics, and important to the design of synthetic active matter \citep{childress1981mechanics,degen2014self,alapan2018soft}. Many cellular organisms exhibit some form of self-propulsion \citep{bray2000cell,lauga2020fluid}, which is usually achieved by flagella or cilia acting on the surrounding fluid \citep{brennen1977fluid}. The motion of microscopic organisms has been widely studied \citep{lauga2009hydrodynamics,marchetti2013hydrodynamics}. At this scale, inertial forces are negligible, i.e. the Reynolds number $Re$ is small. Swimming at large $Re$, where inertia dominates, has also been extensively investigated \citep{becker2015hydrodynamic,maertens2017optimal}. However, the swimming of mesoscale organisms, at $Re$ of order unity, is relatively unexplored \citep{klotsa2019above}. This is due to the simplifications that can be made in Stokes flows $(Re\ll 1)$ and in Euler flows $(Re\gg 1)$, by neglecting inertial and viscous forces respectively, being invalid at intermediate $Re$.

Simplified, or reduced-order, models have been proposed to analyse the locomotion of swimming organisms. A popular example is the squirmer model developed by \citet{lighthill1952squirming} and \citet{blake1971spherical}, wherein self-propulsion is achieved by prescription of a surface velocity, or swimming gait, at the instantaneous surface of the squirmer. Most studies have focused on axisymmetric and impenetrable spherical squirmers, for which the fluid velocity at the surface, and relative to it, can be represented by a modal expansion,
\begin{equation}
    \text{relative surface velocity} =-2\sum_{n=1}^\infty\frac{B_n}{n(n+1)}P_n^1(\cos\theta)\hat{\bm{e}}_\theta,
    \label{eqn:squirmer}
\end{equation}
in which $n$ denotes mode number, $B_n$ the corresponding mode amplitude, $\theta$ the polar angle (and $\hat{\bm{e}}_{\theta}$ its associated unit vector) measured from an arbitrarily chosen `forward' direction along the symmetry axis of the squirmer, and $P_n^m(\cos\theta)$ the associated Legendre polynomials. Note that $n$ odd (even) implies fore-aft anti-symmetric (symmetric) squirming, e.g.  $P_1^1(\cos\theta)=-\sin\theta$, $P_2^1(\cos\theta)=-3\cos\theta\sin\theta$. 

The squirmer model has been  instrumental in examining various aspects of swimming at zero Reynolds number, including hydrodynamic interactions \citep{llopis2010hydrodynamic}; locomotion in viscoelastic fluids \citep{zhu2011locomotion,zhu2012self}; and nutrient uptake \citep{magar2003nutrient,magar2005average}. In that regime, the flow field and squirmer swimming velocity induced by the swimming-gait modal expansion  \eqref{eqn:squirmer} can be obtained by superposing those motions induced by each mode separately. Only the first `dipolar' (fore-aft anti-symmetric) mode contributes to a non-zero swimming velocity \citep{blake1971spherical}. The second `quadrupolar' (fore-aft symmetric) mode contributes a stresslet flow, whose sign distinguishes between `puller' and `pusher' swimmers  \citep{ishikawa2007rheology}: pusher corresponds to negative $B_2$ (relative surface velocity from the equator to the symmetry axis, or poles), and puller to positive $B_2$ (relative surface velocity from the poles to the equator). 

Beyond the Stokes-flow regime, the nonlinear nature of inertia implies that the above superposition principle no longer holds. Previous studies of squirmers at non-zero $Re$ have focused on squirmers whose swimming gait involves only the first two, dipolar and quadrupolar modes in \eqref{eqn:squirmer}. \citet{wang2012inertial} developed an asymptotic expansion through $\mathcal{O}(Re)$ for the swimming speed of a two-mode squirmer at small $Re$, which \citet{khair2014expansions} later extended to $\mathcal{O}(Re^2)$. (In these works, $Re$ is defined based upon the magnitude of the dipolar squirming mode.) \citet{chisholm2016squirmer} performed numerical computations of such a two-mode squirmer for $0< Re< 1000$, bridging the gap between Stokes and Euler flows. They found that, in contrast to the Stokes-flow regime, the swimming speed for a given non-zero value of $B_1$ is affected by the value of $B_2$. For $B_2<0$ (pusher at zero $Re$), increasing $Re$ leads to a monotonic increase in the swimming speed and the axisymmetric flow remains stable to at least $Re=1000$. For $B_2>0$ (puller at zero $Re$), the swimming speed has a non-monotonic dependence on $Re$ and the axisymmetric flow becomes unstable at sufficiently large $Re$. 

\begin{figure}
    \centering
    \includegraphics[width=0.72\linewidth]{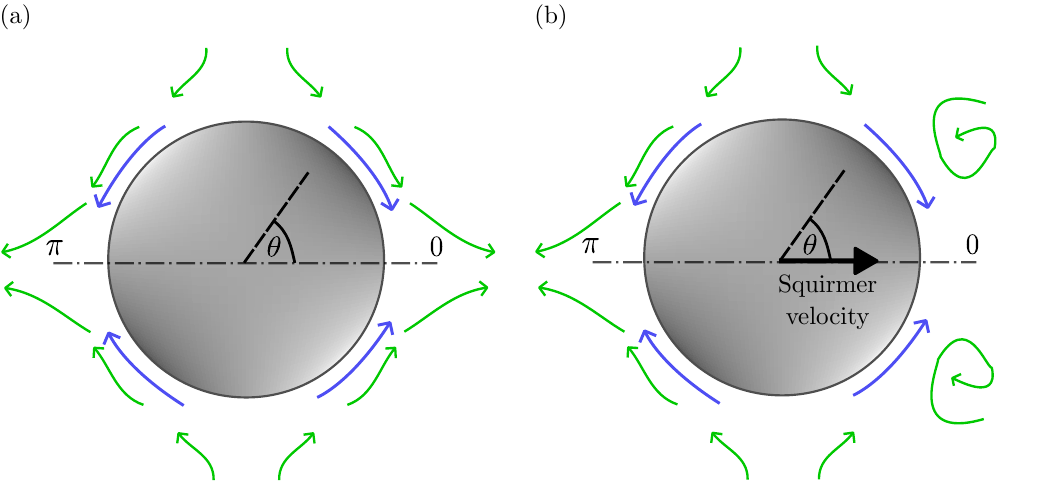}
    \caption{Schematic of spontaneous symmetry breaking of a fore-aft symmetric squirmer. (a) Symmetric steady state, wherein the squirmer is stationary. (b) Symmetry-broken steady state, wherein the squirmer swims. Blue arrows: symmetrically prescribed surface-slip velocity (we show  equator-to-poles squirming as in the case of a quadrupolar pusher).  Green arrows: induced flow in a frame co-moving with the squirmer. }
    \label{fig:squirmer}
    \phantomlabel{a}{fig:base}
    \phantomlabel{b}{fig:sb}
\end{figure}

The scenario considered by \citet{chisholm2016squirmer} was the effect of inertia on a squirmer that is motile at $Re=0$, namely a fore-aft asymmetric squirmer with $B_1\ne0$. It is intuitive that fore-aft asymmetric squirmers that are non-motile at $Re=0$ (i.e. squirmers with $B_1=0$ but $B_n\ne0$ for at least one odd, non-unity value of $n$), generally become motile for $Re>0$, though with their swimming speed vanishing as $Re\searrow0$; this is readily demonstrable by adapting the  small-$Re$ analyses of  \citet{wang2012inertial} and \citet{khair2014expansions}. Here, we ask whether inertia can also enable fore-aft symmetric squirmers to swim via nonlinear symmetry breaking (Fig.~\ref{fig:squirmer}). To this end, we shall numerically study the effects of inertia on quadrupolar squirmers. 

We were led to this question by analogy with recent discoveries of symmetry-breaking locomotion of  isotropically active droplets and particles \citep{michelin2013spontaneous,izri2014self} and Leidenfrost drops \citep{leidenfrostwheels2018}. 
While those examples rely on  nonlinear coupling between hydrodynamics and other physics, inertia alone is well known to result in symmetry breaking in many flow scenarios (e.g. 
the axial asymmetry of wakes 
downstream of a sufficiently fast-moving blunt body). In fact, spontaneous locomotion enabled by inertial symmetry breaking has already been demonstrated for flapping bodies at sufficiently high Reynolds numbers  \citep{vandenberghe2004symmetry,alben2005coherent,vandenberghe2006unidirectional}. 

\section{Problem formulation}
\label{sec:equations}
We consider a spherical, axisymmetric squirmer of radius $a$ that is suspended in a Newtonian, incompressible fluid of density $\rho$ and viscosity $\eta$. Our goal is to study the effects of inertia on a quadrupolar squirmer, whose swimming gait is described by just the second mode in \eqref{eqn:squirmer}, with $B_2$ constant. For the sake of demonstration, we include in the formulation below the possibility to perturb the fore-aft symmetric squirmer forwards/backwards by means of a time-localised dipolar-mode ($n=1$) contribution in \eqref{eqn:squirmer}. 
For simplicity, we assume axial symmetry, and that the squirmer is density matched to the suspending fluid. 

Henceforth, velocities are normalised by $|B_2|$, the pressure and hydrodynamic stress tensor by $\eta |B_2|/a$, and time by $\rho a^2/\eta$. These scales are associated with the Reynolds number $Re=\rho a|B_2|/\eta$. We work in a frame moving with the particle, and employ spherical coordinates $(r,\theta,\phi)$ and associated unit vectors $(\hat{\bm{e}}_r,\hat{\bm{e}}_{\theta},\hat{\bm{e}}_{\phi})$, with origin at the particle centre and $\theta$ the polar angle from the `forward' $\hat{\bm{\imath}}$ direction along the axis of symmetry. 

With these conventions, the fluid flow is governed by the continuity and momentum equations,
\begin{subequations}
\label{NS}
\begin{gather}
    \bm{\nabla\cdot u}=0,\label{eqn:continuity}\\
    \frac{\partial\bm{u}}{\partial t}+\hat{\bm{\imath}}\frac{dU}{dt}+Re \bm{u\cdot\nabla}\bm{u}+\bm{\nabla}p-\nabla^2\bm{u}=\bm{0}, \label{eqn:stokes}
\end{gather}
\end{subequations}
in which $\bm{u}$ and $p$ are the fluid-velocity and pressure fields, $\hat{\bm{\imath}}U$ is the squirmer velocity in the laboratory frame of reference, and $t$ denotes time. The second term on the left-hand side of \eqref{eqn:stokes} represents the fictitious force due to the reference frame's acceleration. With reference to \eqref{eqn:squirmer}, the velocity on the squirmer's boundary is  
\begin{equation}
    \bm{u}=v_s\hat{\bm{e}}_{\theta} ~~~\text{at }r=1, \quad \text{where} \quad v_s=\pm\sin\theta\cos\theta+\lambda(t)\sin\theta.
    \label{eqn:boundary_inner}
\end{equation}
Here, the sign is that of $B_2$ --- thus the plus or minus indicates a puller or pusher, respectively --- and $\lambda(t)$ corresponds to the time-localised dipolar perturbation. (The pusher-puller terminology is based on Stokes-flow theory, where the sign of $B_2$ determines the directionality of the induced force dipole. We have numerically checked that in all cases presented herein inertia does not affect that directionality.) 
Far from the squirmer, 
\begin{equation}
    \bm{u}\to-\hat{\bm{\imath}}U~~~\text{as }r\to\infty.
    \label{eqn:boundary_outer}
\end{equation}
Lastly, the squirmer velocity is coupled to the induced flow via 
Newton's second law,
\begin{equation}
    \frac{4\pi}{3}\frac{dU}{dt}=\hat{\bm{\imath}}\bm{\cdot}\oint_{r=1}\mathcal{N}\bm{\cdot}\hat{\bm{e}}_r\,dS,
    \label{eqn:force}
\end{equation}
wherein $\mathcal{N}=-p\mathcal{I}+\bm{\nabla u}+(\bm{\nabla u})^T$ is the hydrodynamic stress tensor, $\mathcal{I}$ being the identity tensor and the superscript $T$ the tensor transpose, and $dS$ is a dimensionless areal element. 

\section{Methodology}
We next overview our numerical approach to solving the problem formulated in Section \ref{sec:equations}. We shall perform both time-dependent simulations --- initial conditions will be specified later --- and steady-state calculations. Some readers may wish to skip to Section \ref{sec:results}, where we present and discuss our results. 

\subsection{Streamfunction-vorticity formulation}
It is convenient to represent the incompressible flow field $\bm{u}$ in terms of its streamfunction $\psi$, 
\begin{equation}
\bm{u}=\frac{1}{r^2\sin\theta}\frac{\partial \psi}{\partial \theta}\hat{\bm{e}}_r -\frac{1}{r\sin\theta}\frac{\partial \psi}{\partial r}\hat{\bm{e}}_\theta,
\end{equation}
whereby the continuity equation \eqref{eqn:continuity} is trivially  satisfied. The momentum equation  \eqref{eqn:stokes} can then be written 
\begin{equation}
\frac{\partial\bm{\omega}}{\partial t}+Re[\bm{u\cdot\nabla}\bm{\omega}-\bm{\omega\cdot\nabla}\bm{u}]-\nabla^2\bm{\omega}=\bm{0}, \label{eqn:vort_trans}
\end{equation}
wherein the vorticity field $\bm{\omega}=\bm{\nabla}\times\bm{u}$ is azimuthal, i.e. $\bm{\omega}=\hat{\bm{e}}_{\phi}\omega$, the azimuthal component $\omega$ being given in terms of the streamfunction as 
\begin{equation}
\omega=-\frac{1}{r\sin\theta}\left\{\frac{\partial^2\psi}{\partial r^2}+\frac{\sin\theta}{r^2}\frac{\partial}{\partial\theta}\left(\frac{1}{\sin\theta}\frac{\partial\psi}{\partial \theta}\right)\right\}.
\label{eqn:vort_def}
\end{equation}
The boundary and far-field conditions, \eqref{eqn:boundary_inner} and \eqref{eqn:boundary_outer}, become 
\begin{subequations}
\label{axi_conditions}
\begin{gather}
    \psi=0~~~\text{and}~~~\frac{\partial\psi}{\partial r} =-v_s\sin\theta ~~~\text{at }r=1,\\
    \frac{\partial\psi}{\partial r}\sim -Ur\sin^2\theta~~~\text{and}~~~\frac{\partial\omega}{\partial r}\to 0~~~~\text{as }r\to\infty.
    \label{far}
\end{gather}
\end{subequations}
Lastly, the hydrodynamic-force integral in \eqref{eqn:force} can be expressed as 
\begin{gather}
    \hat{\bm{\imath}}\bm{\cdot}\oint_{r=1}\mathcal{N}\bm{\cdot}\hat{\bm{e}}_r\,dS=-\frac{4\pi}{3}\frac{dU}{dt}+
    \pi\int_0^\pi\left[\frac{Re}{2}v_s^2\sin(2\theta)+\left(\frac{\partial(r\omega)}{\partial r}-2\omega\right)\sin^2\theta\right]\,d\theta.\label{eqn:drag}   
\end{gather}
The drag formula \eqref{eqn:drag} has been adapted from \citet{khair2014expansions}. The first term takes into account the fictitious force affecting the pressure due to a non-inertial frame of reference. This term acts as an added mass in \eqref{eqn:force}, effectively doubling the squirmer's inertia. The contribution proportional to $v_s^2$ vanishes for a fore-aft symmetric squirmer, thus in the absence of the dipolar perturbation in the present formulation. 

\subsection{Numerical scheme}
Our method for solving the above formulation as an initial-value problem involves the following two steps. The first consists of solving for the streamfunction and vorticity fields at a given time, given the squirmer velocity and the vorticity field at a previous time. The time derivative in 
\eqref{eqn:vort_trans} is discretised by Backward Euler, 
\begin{equation}
    \bm{\omega}^{(t)}=\bm{\omega}^{(t-\Delta t)}+\Delta t[\nabla^2\bm{\omega}^{(t)}-Re(\bm{u}^{(t)}\bm{\cdot\nabla}\bm{\omega}^{(t)}-\bm{\omega}^{(t)}\bm{\cdot\nabla}\bm{u}^{(t)})],
\end{equation}
where $\Delta t$ is a time step. Eqs.~\eqref{eqn:vort_def} and \eqref{axi_conditions} are written at the present time, except that the previous-time squirmer velocity is used in \eqref{far}. The resulting nonlinear flow problem is solved by a spectral-element method adapted from \citep{chisholm2016squirmer}, which employs the Galerkin method of weighted residuals \citep{karniadakis2005spectral}. The two-dimensional basis set is obtained from a tensor product of one-dimensional Lagrange polynomials of order $N=8$, leading to $(N+1)^2$ degrees of freedom per node, and we make use of Gauss--Lobatto quadrature to integrate over each parametric subdomain. The mesh is generated using Gmsh \citep{geuzaine2009gmsh}, employing a half-ring configuration of inner radius $R_i=1$ and outer radius $R_o=200$. 
The number of nodes is $28^2$, with a radial geometric progression outwards with a factor of $1.25$, while in the polar direction, a factor of $1.1$ is used and the progression direction is towards $\pi/2$. The radial and polar progressions are implemented to handle sharp changes near the squirmer and the wake that occur near the symmetry axis. 
The resulting set of nonlinear algebraic equations is solved using the Newton--Raphson algorithm. For further details about the discretisation, and validation of the method in different scenarios, the reader is referred to \citep{kailasham2022dynamics,cobos2023nonlinear}.  

In the second step, we update the squirmer speed according to \eqref{eqn:force},
\begin{equation}
    U^{(
    t)}=U^{(t-\Delta t)}+\frac{3}{8}\Delta t\int_0^\pi\left[\frac{Re}{2}\left(v_s^{(t)}\right)^2\sin(2\theta)+\left(\frac{\partial}{\partial r}\left(r\omega^{(t)}\right)-2\omega^{(t)}\right)\sin^2\theta\right]d\theta,
\end{equation}
where, consistently with the first step, we discretise the time derivative by Backward Euler. We choose this method since it is unconditionally stable, while maintaining consistency with the first step mentioned above. The two steps are applied iteratively until a given final time is reached ($t=100$ in our simulations). This scheme was validated against \citep{lovalenti1993hydrodynamic} for the time dependent velocity of a sphere subject to a step force at non-zero $Re$. The steady-state problem is solved similarly: the first step remains the same, only that the time-derivative term is dropped from \eqref{eqn:vort_trans}, while the second step is replaced by a secant method to find the value of $U$ that makes the hydrodynamic force vanish. 

\section{Results and discussion}
\label{sec:results}
We have performed time-dependent simulations over the range $0\le Re\le 50$ of puller- and pusher-quadrupolar squirmers. The squirmers are initially at rest, and begin to move in response to a time-localised dipolar perturbation. The flow field at the initial time, $t=0$, is that obtained by the steady-state solver in the absence of the dipolar perturbation and with fore-aft symmetry enforced; we identify that flow as a fore-aft symmetric steady base state of the quadrupolar swimmer (see Fig.~\ref{fig:base}), which constitutes a continuation of the stresslet flow at  $Re=0$ to $Re>0$. The dipolar perturbation is represented by the function $\lambda(t)$ in \eqref{eqn:boundary_inner}, which is chosen to be a Gaussian centred about $t=0.5$, with amplitude $0.1$ and standard deviation $0.1$. (Since the function $\lambda(t)$ is exceedingly small at the initial time, the incompatibility between the perturbed surface velocity and the base-state flow is negligible.)

Fig.~\ref{fig:pusher_time} shows the resulting time evolution of the squirmer velocity in the puller case, for $Re=0,10$ and $50$. The swimming induced by the time-localised dipolar perturbation attenuates for all examined  $Re$, more slowly with increasing  $Re$. For $Re=0$, the velocity nearly traces the perturbation function $\lambda(t)$, as would be expected from Stokes-flow theory (though not precisely owing to a linear inertial effect associated with the rapid variation of the perturbation). For $Re=10$ and $Re=50$, the attenuation of the swimming speed to zero exhibits overshoot; for $Re=50$, the maximum swimming speed in the 
 initial forward-motion phase is actually smaller than that in the later backwards-motion phase. The time evolution of the streamlines is presented on the left-hand side of Fig.~\ref{fig:streams}, for $Re=20$. We note the fore-aft symmetry at the initial and last times (corresponding to the base state), versus the downstream recirculation in both the forward- and backward-motion phases. 

We conclude that for a quadrupolar puller, the symmetric steady base state is stable, at least up to $Re=50$ and under the dipolar perturbations considered. 
\begin{figure}
    \centering
    \includegraphics[width=0.8\linewidth,trim={0.2cm 0.5cm 0 0cm}]{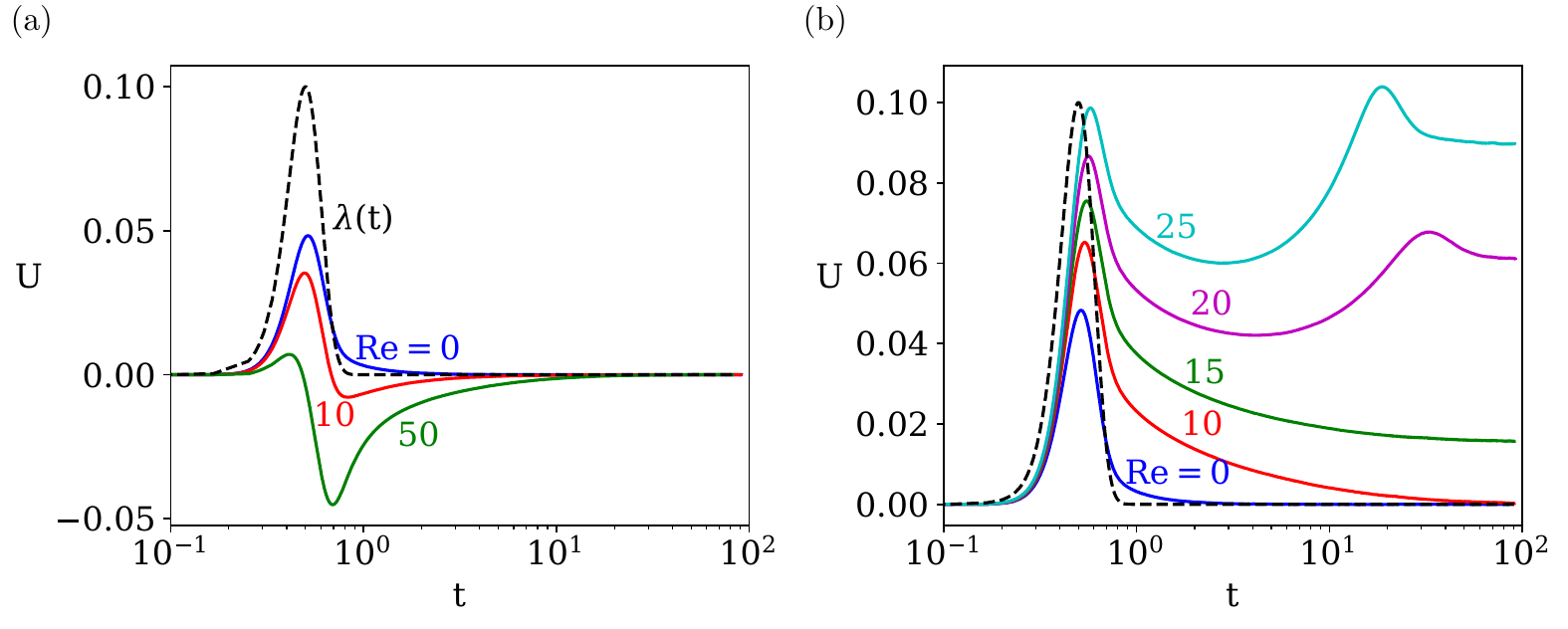}
    \caption{Dimensionless squirmer velocity from time-dependent simulations of initially stationary quadrupolar (a) pullers and (b) pushers, for the indicated $Re$ values. The squirmers are set into motion by means of a time-localised dipolar perturbation represented by the Gaussian function $\lambda(t)$ (cf.~\eqref{eqn:boundary_inner}), depicted by the dashed curves.}
    \label{fig:time_dependent}
    \phantomlabel{a}{fig:pusher_time}
    \phantomlabel{b}{fig:puller_time}
\end{figure}
\begin{figure}
    \centering
    \includegraphics[width=0.73\linewidth]{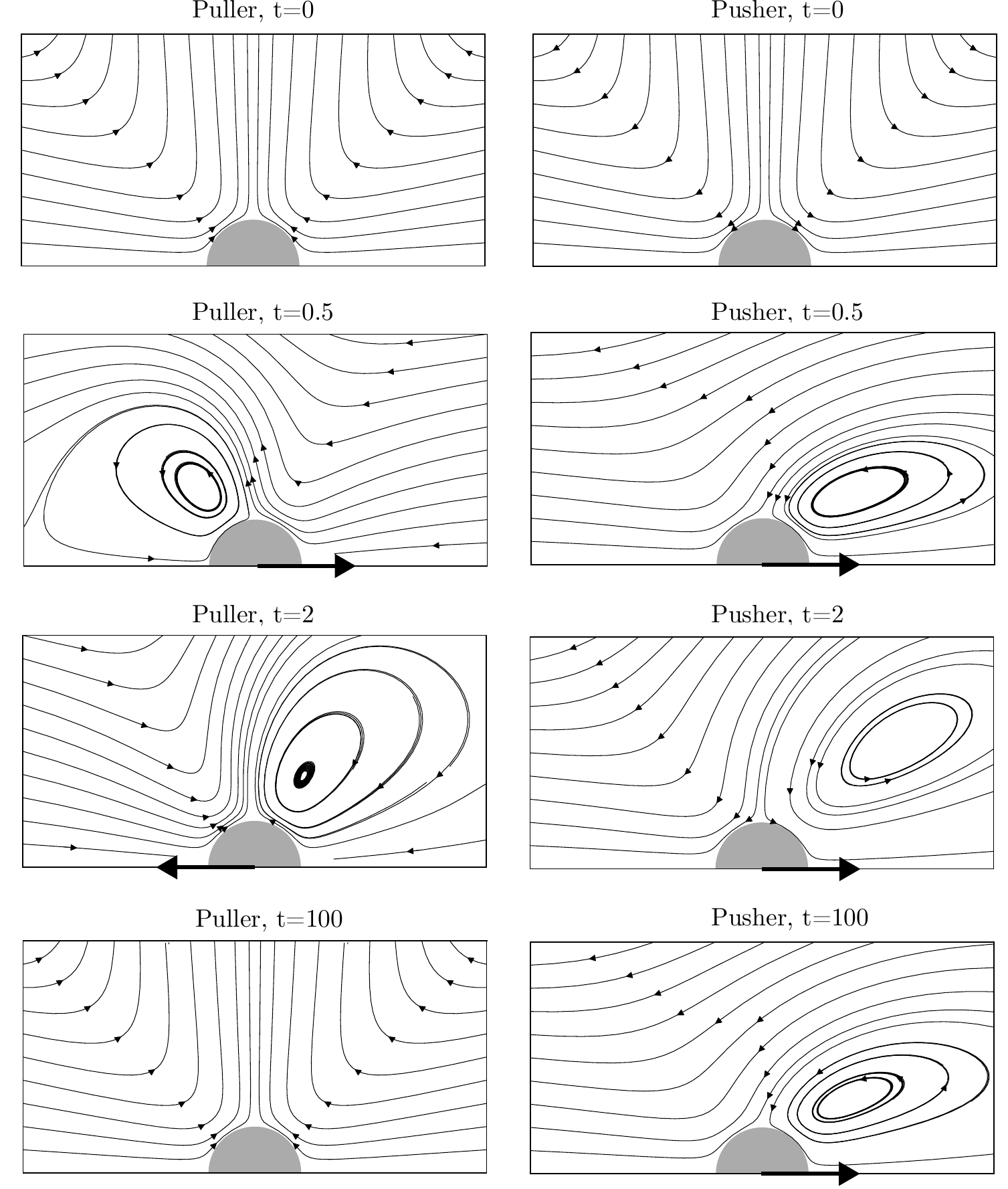}
    \caption{Time evolution of streamlines corresponding to time-dependent simulations as in Fig.~\ref{fig:time_dependent}, for $Re=20$. The dipolar perturbation is maximum at $t=0.5$ and negligible at the other times. The streamlines at $t=100$ are indicative of the steady-state flow patterns.} 
    \label{fig:streams}
\end{figure}

Fig.~\ref{fig:puller_time} shows the time evolution of the squirmer velocity in the pusher case, for $Re=0,10,15,20$ and $25$. The response of pushers to the dipolar perturbation is seen to dramatically differ from that of pullers. Immediately following the attenuation of the perturbation, the swimming velocity attenuates as well, more slowly with increasing $Re$ and without reversing direction as in the puller case. For $Re\leq 10$, this attenuation persists such that the swimming speed vanishes at long times. In contrast, for $Re\geq15$, the swimming speed approaches a non-zero value, which increases with $Re$; for $Re\geq20$, the approach to steady-state swimming is non-monotonic. The time evolution of the streamlines is presented on the right-hand side of Fig.~\ref{fig:streams}, for $Re=20$. We note the upstream recirculation generated by the squirmer's motion, which contrasts the downstream recirculation observed in the puller case. 

We conclude that the symmetric steady base state of a  quadrupolar pusher is unstable beyond a critical Reynolds number. Following a time-localised dipolar perturbation, the dynamics are seen to approach a symmetry-broken steady state where the squirmer exhibits self-sustained locomotion and the flow around the squirmer is fore-aft asymmetric (Fig.~\ref{fig:sb}). While the spontaneous locomotion is `forward' in our time-dependent simulations, it is clear from the symmetry of the problem that a mirror-reflected swimming state could be excited by flipping the sign of the dipolar perturbation. 

\begin{figure}
    \centering    \includegraphics[width=0.73\linewidth,trim={0 0.5cm 0 0}]{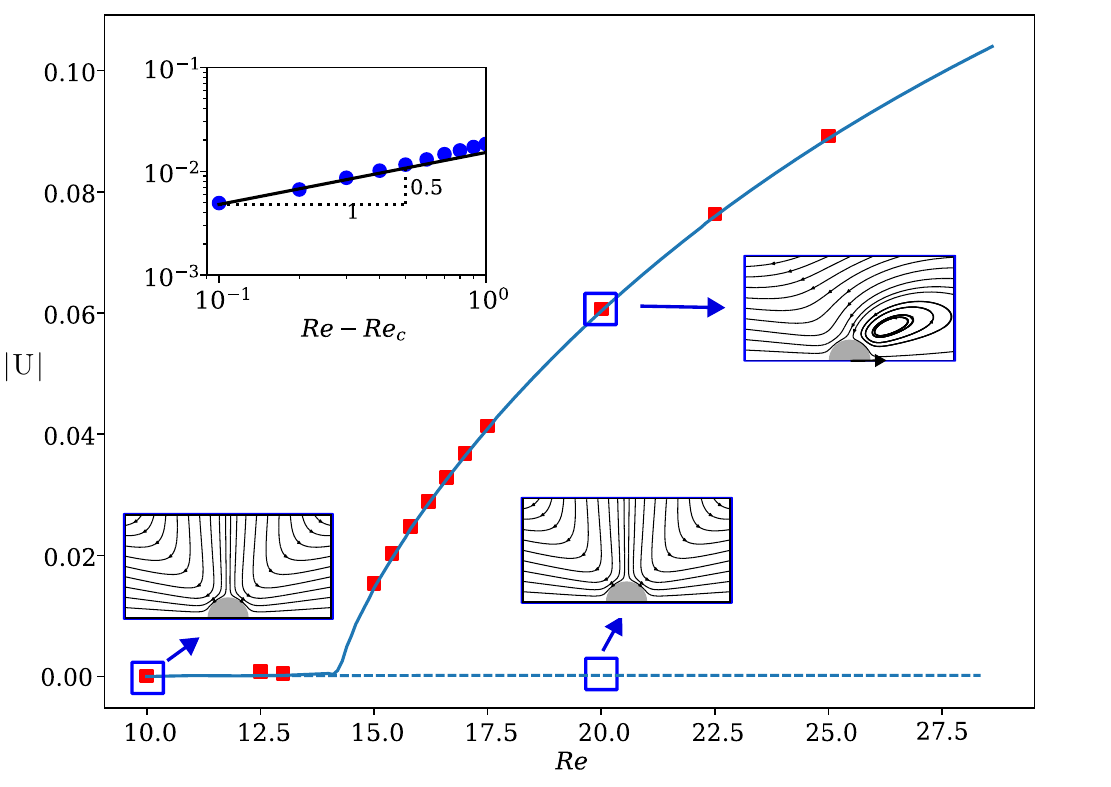}
    \caption{Steady swimming velocity $U$ vs $Re$ for a quadrupolar-pusher squirmer. Blue curves: steady-state computations employing a fore-aft asymmetric (solid) and symmetric (dashed) initial guess. Red squares: final velocity in the time-dependent simulations. The insets show the streamlines at the indicated $Re$ and confirm the $|U|\propto (Re-Re_c)^{1/2}$ behaviour near the swimming threshold, which is canonical of a pitchfork bifurcation.}
    \label{fig:steady}    
\end{figure}

From Fig.~\ref{fig:puller_time}, we conclude that the critical Reynolds number for a quadrupolar pusher (at least under dipolar perturbations), say $Re_c$,  lies between $10$ and $15$. However, pinning down the specific value of $Re_c$ with time-dependent computations is prohibitively expensive because the steady swimming speed is asymptotically small as $Re\searrow Re_c$. Therefore, we turn to steady-state calculations. The resulting bifurcation diagram of $U$ as a function of $Re$ is shown in Fig.~\ref{fig:steady}. The critical Reynolds number is found to be $Re_c\approx 14.3$, with the steady swimming speed beyond the bifurcation monotonically growing with $Re$. As shown in the inset, $|U|\propto \sqrt{Re-Re_c}$ for  $Re$ near $Re_c$, in agreement with the canonical scaling expected for a pitchfork bifurcation. The figure also depicts sample streamlines of the symmetric steady state at $Re=10$ and $Re=20$, and the symmetry-broken swimming state at $Re=20$. 

It is interesting to contrast the observed scaling for $U$ near the bifurcation threshold with that for an isotropic  autophoretic particle at $Re=0$ \citep{michelin2013spontaneous}. In the autophoretic problem, an instability leading to spontaneous locomotion is observed at a sufficiently large dimensionless rate of solute emission, or intrinsic  P\'eclet number $Pe$. Instead of the canonical scaling for a pitchfork bifurcation, as found herein, the particle speed is found to obey $|U|\propto Pe-Pe_c$, for $Pe$ near its critical value $Pe_c$ \citep{morozov2019self,saha2021isotropically,kailasham2022dynamics,schnitzer2023weakly}. This linear scaling can be traced to the fact that the base state for a spherical autophoretic particle involves no flow, hence the effective P\'eclet number (indicating the true ratio of advection to diffusion) near the bifurcation is actually small, such that nonlinear advection is to leading-order negligible except at large distances from the particle. Nonetheless, an analogy can be drawn with the case of a fore-aft symmetric, yet non-isotropic autophoretic particle, such as the homogeneous elliptical particles recently studied by \cite{zhu2023self}. In such cases, the base state involves a non-trivial fore-aft symmetric flow, as herein, and indeed the bifurcation is canonical.

\section{Concluding remarks}\label{sec:conclusion}
We have shown via axisymmetric numerical simulations of the Navier--Stokes equations that quadrupolar-pusher squirmers, which possess axial and fore-aft symmetry, are  capable of self-sustained locomotion above a moderate critical Reynolds number, $Re_c\approx 14.3$. Beyond that threshold, the steady swimming speed monotonically increases with $Re$ over the range of $Re$ examined, initially like $(Re-Re_c)^{1/2}$. Our simulations have further demonstrated that the symmetric base state, in which the squirmer is stationary, becomes unstable above the swimming threshold; in particular, when that state is disturbed by a time-localised dipolar perturbation, the squirmer relaxes towards steady swimming. These results together suggest that the spontaneous swimming emerges through a supercritical pitchfork bifurcation. 

As far as we are aware, this paper is the first to suggest, let alone demonstrate, the possibility of spontaneous squirmer locomotion arising from an inertial symmetry breaking. As such, our findings give rise to many intriguing questions, which call for more extensive numerical investigations, as well as theoretical analyses:
\begin{enumerate}
\item Besides quadrupolar pushers, it is clear that other, more general fore-aft symmetric squirmers are also capable of spontaneous locomotion. How does the critical Reynolds number and swimming speed depend on the combination of even modes in the modal expansion \eqref{eqn:squirmer}? Computationally optimising the swimming gait with respect to some appropriate cost function, such as the swimming efficiency used at low Reynolds numbers \citep{lauga2009hydrodynamics}, would facilitate a comparison between spontaneous and conventional swimming at moderate Reynolds numbers.   
\item Our simulations were limited to  moderate $Re$. Up to what $Re$ can a quadrupolar pusher sustain stable locomotion, with a speed monotonically increasing with $Re$? A maximum with respect to the bifurcation parameter and arrest of the swimming for values of that   parameter sufficiently away from the onset of symmetry breaking were found for active droplets and particles \citep{michelin2013spontaneous,izri2014self}, and Leidenfrost drops \citep{leidenfrostwheels2018}. 
\item In our simulations, the squirmer motion is collinear and the flow field is axisymmetric. For what range of $Re$, if at all, do the spontaneous-swimming states identified here remain stable under general three-dimensional perturbations? Even if the axisymmetric swimming states are unstable, they may have stable variants featuring axial-symmetry breaking, in addition to the fore-aft-symmetry breaking. Furthermore, can some fore-aft symmetric squirmers also sustain swimming normal to their axis of symmetry?
\item How would the spontaneous swimming of a symmetric squirmer be affected by weak asymmetric perturbations, such as owing to a constant fore-aft asymmetric swimming gait perturbation, gravity, or interactions with other swimmers and with boundaries? Given the nature of spontaneous swimming, we generally expect such perturbations to significantly affect the squirmer's dynamics---not directly, but rather by slowly redirecting the motion. For example, studies of perturbed active colloids \citep{saha2021isotropically,kailasham2022dynamics,li2022dynamics,schnitzer2023weakly,peng2023weakly,zhu2023self} have demonstrated that such perturbations can introduce asymmetry between forward and backward locomotion manifested in both speed and stability, and in some cases also promote meandering or curvilinear spontaneous motion. 
\end{enumerate}

We conclude by noting two potential broader implications of this work. First, given the biological context of the squirmer model, the uncovered inertial symmetry breaking presents a possible mechanism for locomotion of moderate-Reynolds-number organisms. Second, our findings may lead to new design strategies for robotic swimmers that spontaneously undergo locomotion depending on the flow conditions they encounter, e.g. the viscosity of the surrounding fluid.

\textbf{Acknowledgement.} This work was supported by the National Science Foundation through grant number CBET 2002120. OS acknowledges the support of the Leverhulme Trust through Research Project Grant RPG-2021-161.

%
%

\bibliographystyle{jfm}
\bibliography{jfm}
\end{document}